\documentclass[aps,pre,twocolumn,showpacs,floatfix,preprintnumbers,superscriptaddress,amsmath,amssymb]{revtex4-1}
\usepackage{amsmath}
\usepackage{amssymb}
\usepackage{amsfonts}
\usepackage{graphicx}
\usepackage{epsfig}
\usepackage{epstopdf}
\usepackage{dcolumn}
\usepackage{bm}
\usepackage{array}
\usepackage{mathtools}
\usepackage{url}
\usepackage[colorlinks,bookmarks=false,citecolor=blue,linkcolor=red,urlcolor=blue]{hyperref}

\begin{document}

\title{Study of loss dynamics of strontium in a magneto-optical trap}
\author{Chetan Vishwakarma}
\affiliation{Department of Physics, Indian Institute of Science Education and Research, Pune 411008, Maharashtra, India}
\author{Kushal Patel}
\affiliation{Department of Physics, Indian Institute of Science Education and Research, Pune 411008, Maharashtra, India}
\author{Jay Mangaonkar}
\affiliation{Department of Physics, Indian Institute of Science Education and Research, Pune 411008, Maharashtra, India}
\author{Jamie L. MacLennan}
\affiliation{Department of Physics, University of Michigan, Ann Arbor, Michigan 48109, USA}
\author{Korak Biswas}
\affiliation{Department of Physics, Indian Institute of Science Education and Research, Pune 411008, Maharashtra, India}
\author{Umakant D. Rapol}
\email {Electronic mail: umakant.rapol@iiserpune.ac.in}
\affiliation{Department of Physics, Indian Institute of Science Education and Research, Pune 411008, Maharashtra, India}

\begin{abstract}
Collisions with background atoms are known to induce a significant shift in the frequency of state-of-the-art optical atomic clocks and contribute to state decoherence in cold atom experiments. The effects of these collisions can be quantified by measuring their cross sections. We experimentally measured the collision cross section between $^{88}$Sr$-$N$_{2}$ in a Magneto-Optical Trap (MOT). The measurement was carried out by monitoring the atom number loss rate as a function of background pressure of N$_{2}$ and the cross section thus obtained was 8.1(4)$\times 10^{-18}$ m$^{2}$. The measured collision cross section has been utilized for the determination of C$_{6}$ coefficient of the ground state (${^1S}_0$) of $^{88}$Sr atom, which can be useful to estimate the relative frequency shift in the clock transition. We also estimate the loss rate induced by the combined effect of the decay of atoms in the long-lived ${^3P}_0$ state and temperature-induced atomic losses from the capture volume of the MOT. We find that the contribution due to the latter is dominant in comparison to the other atomic loss channels and must be included in the studies that rely on the total loss rate measurement. 
\end{abstract}

\maketitle
\section{Introduction}
The invention of optical frequency comb \cite{RevModPhys.75.325} in conjunction with laser cooling techniques has boosted efforts towards using optical transitions as a universal time standard. Among various architectures \cite{RevModPhys.87.637}, neutral atom based platforms have become popular in realizing the next generation optical clocks which hold promise to revolutionize global timekeeping, precision sensing and probing the stability of fundamental constants \cite{RevModPhys.87.637,TINO2007159,Lea_2007}. The clock accuracy depends on accurately quantifying various systematic shifts in the clock transition frequency; further progress will require reducing their respective uncertainties.\

Frequency shifts due to Background Gas Collisions (BGCs) are currently one of the largest sources of uncertainty in many of the best atomic clocks of various types \cite{Bothwell2019,Nicholson2015,McGrew2018,PhysRevLett.123.033201,PhysRevA.96.022704,PhysRevLett.116.063001,PhysRevA.87.023806,PhysRevLett.104.070802,Cao2017}. This has motivated a number of theoretical \cite{PhysRevLett.110.180802,Cybulski2019,Davis2019,Vutha2018,PhysRevA.100.033419} and experimental \cite{Alves2019,McGrew2018} efforts to study these shifts. In particular, in Ref. \cite{PhysRevLett.110.180802} a model is developed relating the shifts to dispersion coefficients (and thus to collisional cross sections), facilitating the estimation of shifts without directly probing the clock transitions.\

One of the leading candidates in neutral-atom based clocks is the Sr optical lattice clock, with a current performance of 2.0$-$2.1$ \times 10^{-18}$ uncertainty in the two most-accurate ones \cite{Bothwell2019,Nicholson2015}. In these, BGC-induced shifts are the third largest uncertainty, contributing 4$-$6$\times 10^{-19}$ to the uncertainty budget. Recent work \cite{Alves2019} had enabled a reduction in this contribution in the former clock \cite{Bothwell2019} by a measurement of the frequency shift due to collisions with H$_{2}$, with ongoing studies to investigate the effect of other species present in the background gas. A theoretical estimate for the shift due to Sr$-$H$_{2}$ collisions was also recently made \cite{Cybulski2019,AbdelHafiz2019}.\

While previous studies on BGC-induced shifts have mainly focused on H$_{2}$ because it is typically the dominant species present in the vacuum chamber, it is anticipated that as the clock uncertainties continue to improve, it may become necessary to account for the contribution of other background gas species such as N$_{2}$. Dispersion coefficients have been estimated for collisions of N$_{2}$ with alkali, noble, and various molecular gases and collisional cross sections have been reported for Rb$-$N$_{2}$ \cite{PhysRevA.64.023402,Booth2019}, Ne*$-$N$_{2}$\cite{PhysRevA.78.042712,Glover2010}, Na$-$N$_{2}$\cite{Prentiss1988} and Ar$-$N$_{2}$\cite{Brunetti1983}, but so far Sr$-$N$_{2}$ collisional properties have not been investigated.\ 

The traditional way to determine the collision cross section is to employ crossed beam technique \cite{doi:10.1063/1.473420,Kau_1996} where, two collimated atomic/molecular beams are generated and made to intersect in a well-defined interaction region. Uncertainties in the number of target atoms and the volume of the intersection region are two major sources of uncertainty in the measurement. During an experiment, these are the two major sources of errors. Since the invention of the technique of laser cooling, an alternative way to measure the cross section is the atomic loss rate from the MOT, magnetic trap (MT), or the Optical Dipole Trap (ODT). Collision cross sections determined in this manner have been shown to be more accurate with respect to the beam-based method \cite{PhysRevA.78.042712}. There have been extensive experimental studies performed in the same spirit using Rubidium \cite{PhysRevA.85.033420,doi:10.1063/1.4928154,PhysRevA.64.023402}, Ytterbium \cite{Rapol2004}, Cesium \cite{Yuan:13}, Neon \cite{doi:10.1063/1.2754444,PhysRevA.78.042712} etc.\    

In this article, we study the dynamics of a Sr MOT, by observing the loading  and the loss rate under various conditions and repumping schemes. Using this data, we also report the first experimental determination of $^{88}$Sr$-$N$_{2}$ collision cross section of the ${^1S}_0$ ground state by measuring the loss rate of atoms from $^{88}$Sr MOT operating with the first stage cooling transition. The collision cross section is determined by injecting nitrogen (N$_{2}$) inside the vacuum chamber in a controlled manner and studying the atomic loss rate at different background pressures. This measurement is helpful in evaluation of systematic shift in the fractional frequency uncertainty in the clock operation. The presence of intermediate metastable states which the atoms in the exited state (${^1P}_1$) of the cooling transition can decay into, complicates the loss rate dynamics. To study this, we consider the various loss mechanisms in MOT and evaluate their contributions. In our experiment, the total atomic loss rate depends on (i) the collision with untrapped background atoms/molecules, (ii) the decay of atoms into the triplet state ${^3}P_0$, and (iii) the temperature of the atomic cloud.\  

We present a model to accommodate the above loss channels. In order to check the consistency of the model for loss rate determination, we have performed the experiment in presence of two different repumping schemes utilizing (a) 707.2 nm laser connecting $5s\ 5p\ {^3P}_2\longrightarrow 5s\ 6s\ {^3S}_1$ and (b) with 481.3 nm laser addressing $5s\ 5p\ {^3P}_2\longrightarrow 5p^{2}\ {^3P}_2^{'}$ transition. For even isotopes, the transition from ${^3P}_0$ to the ground state ${^1S}_0$ is forbidden under spin selection rule and thus contributes to the total atomic loss rate. Due to the existence of multiple decay channels from the excited state, the atoms spend significant amount of time in the states which are unresponsive to the first stage cooling laser. These atoms contribute to the total loss rate, if the collective lifetime of these states is larger than the time required for the atoms to escape the capture region.\ 

This loss channel is proportional to the atomic cloud temperature and is the dominant channel for the atomic species under consideration. It is important to consider this loss channel for the atomic species which have level structure similar to strontium. The contribution of this term, along with losses induced by decay into the triplet state, is determined by studying the loss rate as a function of MOT beam intensity.\

\section{Theoretical background}

\begin{figure}[ht]
	\centering
	\includegraphics[width=0.98\linewidth]{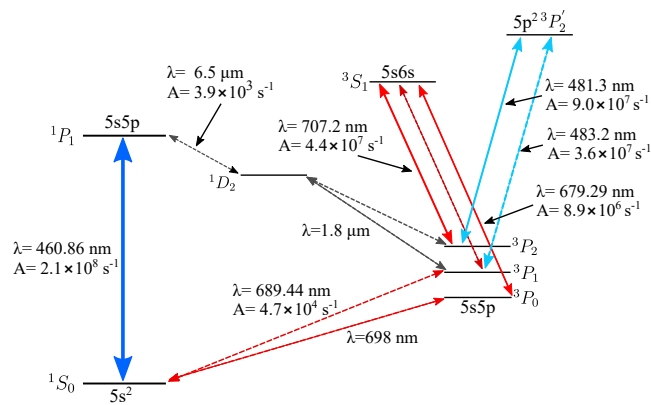} 
	\caption{The relevant low-lying energy level diagram of strontium ($^{88}$Sr) atoms. The wavelength ($\lambda$) and the decay rate (A) has been shown along with the transition.}
\label{fig1}
\end{figure}

The experiment is performed with the most abundant even isotope of strontium ($^{88}$Sr). Fig. \ref{fig1}, shows the low-lying energy level diagram of $^{88}$Sr atom. It has two cooling transitions $5s^2\ {^1S}_0\longrightarrow 5s\ 5p\ {^1P}_1$ and $5s^2\ {^1S}_0\longrightarrow 5s\ 5p\ {^3P}_1$ of wavelengths 460.8 nm and 689.4 nm respectively. The singlet to triplet transition (${^1S}_0\longrightarrow {^3P}_1$) has a long lifetime of 21 $\mu$s and linewidth of 2$\pi \times$7.5 kHz. This transition can be used for the second stage of cooling and has extremely low Doppler limited temperature of  $\sim$180 nK. The other cooling transition connects ${^1S}_0 \longrightarrow {^1P}_1$. This transition has a short lifetime of $\sim$5 ns ($\Gamma$ = 2$\pi\times$32 MHz) and is well suited for the first stage of cooling and trapping. The larger linewidth results in a higher Doppler limited temperature of $\sim$770 $\mu$K; further, the transition itself is not completely closed. The atoms in the excited state ${^1P}_1$ decays to ${^1D}_2$ state with a branching ratio of 1$\colon2\times 10^{-5}$. Atoms in ${^1D}_2$ further decay to  ${^3P}_2$ and  ${^3P}_1$ states with a branching ratio of 1$\colon$2, respectively. The transition $\ {^3P}_2 \longrightarrow {^1S}_0$ state is doubly forbidden; thus, the atoms once decayed to ${^3P}_2$  are lost from the cooling cycle. There are several repumping schemes to revive this loss  \cite{PhysRevA.90.022512,PhysRevA.71.061403,Moriya_2018,Mickelson_2009,PhysRevA.99.033422}. The most popular of them employ two lasers operating at 707.2 nm and 679.3 nm. The decay rates and wavelengths of the transitions relevant for the present study are shown in Fig. \ref{fig1}.\

The studies reported here are based on the change in the decay rate of atoms from the MOT as a function of background gas density and the trapping beam intensity. At room temperature, the untrapped background particles have sufficient kinetic energy to knock the atoms out of the trap. These collisions limit the atomic lifetime in the trap and are directly proportional to the background pressure inside the vacuum chamber. The loading of atoms in the MOT is a dynamic process, and a steady-state population is achieved on attaining equilibrium between the loading and the loss rate. The rate equation can be written as:
\begin{equation}
\frac{dN}{dt}= R-\frac{N}{\tau}-\beta N^{2}
\label{eq1}
\end{equation}
In the above expression, $N$ is the number of atoms at a given instant of time `$t$', $R$ is the loading rate,  $1/\tau$ is the linear loss rate of atoms, and $\beta$ represents the non-linear loss rate due to intra-trap collisions. The experiments reported in this article have been performed in the linear regime, and thus the term which is quadratic in $N$ is neglected \cite{PhysRevA.59.1216} for further analysis. In the low-density regime, the solution to Eq. \ref{eq1} displays an exponential growth in the number of atoms in MOT and can be written as $N = N_{s} \left[1 - \exp{(\frac{-t}{\tau})}\right]$, where, $N_{s} = R \tau$.\

The linear loss rate in Eq. \ref{eq1} can be attributed to several factors, namely, (a) collision with the background N$_{2}$, (b) collision with background species other than N$_{2}$, (c) decay of atoms into the metastable states and (d) escape of atoms out of the capture region. By combining the factors mentioned above, the total linear loss rate can be written as:
\begin{equation}
\begin{aligned}
\frac{1}{\tau} = \quad & n_{N_{2}}\sigma_{N_{2}}v_{N_{2}} + \Gamma_{b} + \alpha_{0} f + \\
 & \epsilon\gamma_{t}\frac{A_{\ {^1P}_1\rightarrow\ ^1D_2}\ f}{A_{\ {^1D}_2\rightarrow\ ^3P_2}+A_{\ {^1D}_2\rightarrow\ ^3P_1}} 
\end{aligned}
\label{eq2}
\end{equation}
Here, $n$ is the density of the background particle, $\sigma$ is the collision cross section, and $v$ is the average velocity of the particle under consideration. The subscript $N_{2}$ denotes the background species used for the current experiment. The second term, $\Gamma_{b}$ in the above equation accounts for the loss induced by the background species other than N$_{2}$. The term $\alpha_{0}f$ in the above equation represents the loss of atoms due to branching into the metastable state. In our experiment (considering the cooling laser at 460.8 nm and repumping lasers at 707.2 nm and 481.3 nm for two different schemes), such a loss corresponds to the effective decay of atoms into the long-lived state $^3P_0$ and is denoted by $\alpha_{0}f$. This loss channel can be explained in the following way. The atoms in excited state ${^1P}_1$ decay to the intermediate state ${^1D}_2$ with the rate $A_{\ {^1P}_1\rightarrow\ {^1D}_2}$. These ${^1D}_2$ atoms can further decay to ${^3P}_2$ state with the branching ratio $B_{\ {^1D}_2\rightarrow\ {^3P}_2}$. A fraction of atoms is excited into the $5s\ 6s$ ${^3S}_1$ (5p$^2$ ${^3P}_2^{'}$) state by using the repumping laser at $\lambda = 707.2$ nm ($\lambda = 481.3$ nm). These atoms can make a transition from   ${^3S}_1$ (${^3P}_2^{'}$) to ${^3P}_0$ state with the branching ratio $B_{\ {^3S}_1({^3P}_2^{'})\rightarrow\ {^3P}_0}$. For the isotope under consideration, the transition from the triplet state ${^3P}_0$ to the ground state ${^1S}_0$ is forbidden under spin selection rule. Consequently, the process of populating the ${^3P}_0$ state is a loss channel. As the fraction of atoms in the excited state ${^1P}_1$ depends on the power of MOT beams, this loss rate is denoted by $1/\tau_{power}$. Thus, the third term of Eq. \ref{eq2} can be expanded as \cite{Xu:03}:
\begin{equation}
\begin{aligned}
\frac{1}{\tau_{power}}=\alpha_{0} f=(fA_{\ {{^1P}_1\rightarrow\ {^1D}_2}})\times  (B_{\ {{^1D}_2\rightarrow\ {^3P}_2}}) \times & \\ 
(f^{'}B_{\ {{^3S}_1({^3P}_2^{'})\rightarrow\ {^3P}_0}}) 
\end{aligned}
\label{eq3} 
\end{equation}
Here, $\textit{f}$ and ${f^{'}}$ is the fraction of atoms in the excited state ${^1P}_1$ and ${^3S}_1({^3P}_2^{'})$ respectively. The general expression for these fractions is be given as:
\begin{equation}
f=\frac{1}{2}\frac{I/I_{0}}{1+I/I_{0}+(2\Delta/\Gamma)^2}
\label{eq4}
\end{equation}
where, $\textit{I}$ is the total intensity, $I_{0}$ is the saturation intensity of the transition, $\Delta$ is detuning of the laser beam, and $\Gamma$ is natural linewidth of the transition under consideration.\

The last term $\left( \epsilon \gamma_{t}\frac{A_{\ {{^1P}_1\rightarrow\ {^1D}_2}}f}{A_{\ {{^1D}_2\rightarrow\ {^3P}_2}}+{A_{\ {{^1D}_2\rightarrow\ {^3P}_1}}}}\right)$ in Eq. \ref{eq2} represents the atomic loss due to escape from the capture region and will be denoted by $1/\tau_{temp}$ for further analysis in this manuscript. The detailed methodology for the calculation of this loss channel is given in Ref. \cite{Kurosu_1992}. The term  ${A_{\ {{^1D}_2\rightarrow\ {^3P}_2}({^3P}_1)}}$ denotes the decay rate from the intermediate state  ${^1D}_2$ to the triplet state  ${^3P}_2({^3P}_1)$. This loss channel can be understood as follows. A fraction of atoms from the excited state ${^1P}_1$ decay to the intermediate state ${^1D}_2$, which further connects to two triplet states ${^3P}_2$ and ${^3P}_1$ as shown in Fig. \ref{fig1}. The atoms in ${^3P}_1$ state decay to the ground state in 21 $\mu$s. On the other hand, the atoms that have decayed to ${^3P}_2$ state can be brought back to the main cooling transition by using repumping lasers. Considering all the cascade channels, it takes $\sim$1 ms ($\equiv\gamma_{t}^{-1}$) for the atoms in ${^1D}_2$ to return to the ground state via ${^3P}_2$ and ${^3P}_1$ state. During this process, the atoms are not responsive to the main cooling laser operating at 460.8 nm and therefore have a finite probability of escaping from the capture region of MOT. This probability ($\epsilon$) is determined by the diameter of the MOT beams and the average temperature of the atomic ensemble \cite{Kurosu_1992}:  
\begin{equation}
\epsilon \sim \int_{0}^{\infty} \left(\frac{1}{2{\pi} v^{2}_{0}}\right)^\frac{3}{2} 4\pi v^{2} exp\left(\frac{-v^{2}}{2v^{2}_{0}}\right)exp\left(\frac{-R\gamma_{t}}{v}\right)dv
\label{eq5}
\end{equation}
Here, $v_{0}$ and $R$ denote the root mean square velocity of the atoms and the radius of the trapping beam, respectively.\ 

After the identification of various loss mechanisms, the individual contributions have been evaluated by operating the MOT at different experimental conditions. The measurement of the total loss rate as a function of power in the trapping beams and density of the background N$_{2}$ gas provides relevant information regarding the collision cross section and losses due to branching into the long-lived state $^{3}P_{0}$. 

\section{Experimental Details}
The experiments are performed using the blue MOT of $^{88}$Sr atoms formed inside a cuboidal quartz cell having dimensions of 38 mm$\times$38 mm$\times$160 mm. We use an effusive source \cite{doi:10.1063/1.5067306} of atoms, which is resistively heated to 600$^\circ$C.  These atoms are then slowed down by a zero-crossing Zeeman slower to load the MOT. The MOT laser beams are prepared following the standard $\sigma^{+} - \sigma^{-}$ configuration with retro-reflection geometry. The light for the first stage laser cooling ($5s^2\ {^1S}_0\longrightarrow 5s\ 5p\ {^1P}_1$ , $I_s = 42.7 \ $mW/cm$^2$) is generated using a homemade cavity-enhanced optical frequency doubler. The frequency doubler uses a periodically poled KTP crystal for Second Harmonic Generation (SHG). This cavity is injected with an input laser ($\lambda =  922$ nm, power = 700 mW) light from a commercial tapered amplifier and generates $\sim$200 mW of blue light. The seed light for the amplifier is generated using a grating-stabilized diode laser in Littrow configuration. The frequency of the laser is stabilized to the cooling transition by performing atomic beam spectroscopy \cite{doi:10.1063/1.4977593}. The magnetic field for the MOT is generated using a pair of anti-Helmholtz coils with an axial field gradient of $\sim$48 Gauss/cm. During the loading process, detuning of the MOT beams is kept at $ -$33 MHz.

\begin{figure}[ht]
	\centering
	\includegraphics[width=0.95\linewidth]{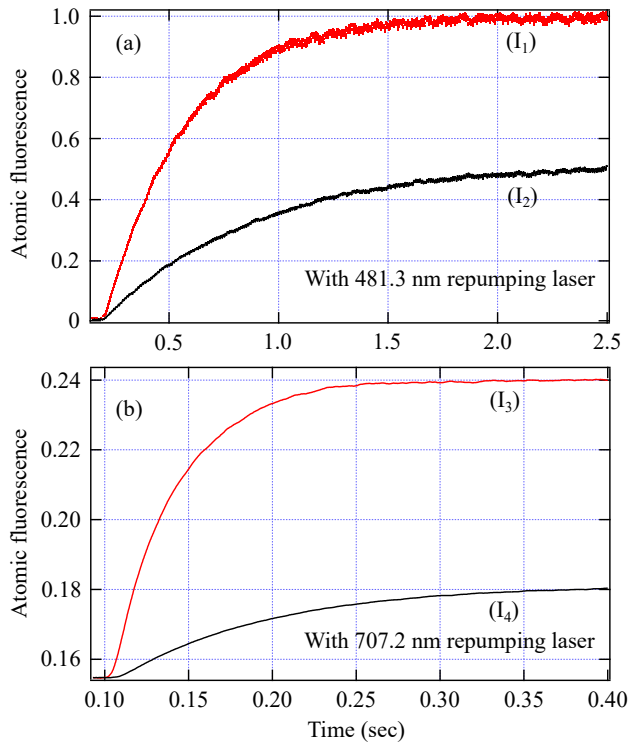} 
	\caption{Normalized fluorescence of $^{88}$Sr MOT in the presence of repumping laser operating at (a) 481.3 nm and (b) 707.2 nm for different trapping intensities, I$_{1}$ = 46.4 mW/cm$^{2}$, I$_{2}$ = 26.4 mW/cm$^{}$, I$_{3}$ = 96.8 mW/cm$^{2}$ and I$_{4}$ = 31 mW/cm$^{2}$. All other parameters are kept constant. Note that the loading rates of the MOT for the two repumping schemes is significantly different.}
	\label{fig2}
\end{figure}

The experiments have been conducted in two different configurations. In the first set, each of the MOT beams have 1/$e^{2}$ diameter of 10 mm and maximum intensity of $\sim$16 mW/cm$^2$. The decay of atoms into the metastable state ${^3P}_2$ is suppressed using a repumping laser of wavelength 707.2 nm operating on the  $5s\ 5p\ {^3P}_2\longrightarrow 5s\ 6s\ {^3S}_1$  transition. This transition has a saturation intensity $I_s = 3.4 \ $mW/cm$^2$. For the second set of experiments, the 1/$e^{2}$ diameter of the MOT beams has been increased to 15 mm. The maximum intensity of each beam in this configuration is $\sim$8 mW/cm$^2$. For this set, we used the second repumping scheme with wavelength 481.3 nm ($I_s = 0.028 \ $mW/cm$^2$) \cite{PhysRevA.99.033422}. This laser connects the transition $5s\ 5p\ {^3P}_2\longrightarrow 5p^{2}\ {^3P}_2^{'}$ as shown in Fig. \ref{fig1}. The frequency of both the repumping lasers is locked using a digital PID controller which receives  feedback of the laser frequency via a commercial wave-meter ({\em High Finesse}, model: WSU-30) \cite{doi:10.1063/1.5025537}. The number of trapped atoms in MOT is estimated by collecting the fluorescence of the atoms on a calibrated photomultiplier tube ({\em Hamamatsu}, model H9307-02).\

The determination of the collision cross section is based on the change in the MOT loading rate on the introduction of gas species inside the vacuum chamber. The vacuum system for the current experiment consists of two sections -- the oven region and the quartz cell connected to the Zeeman slower tube. These two regions are separated by a differential pumping tube. The pressure in these sections is maintained by two independent ion pumps  (capacity $\sim$55 {\em l} s$^{-1}$). Also, a Titanium Sublimation Pump (TSP) is connected to the science chamber (near the quartz cell) for maintaining the desired low pressure.  The base pressure inside the quartz cell is less than 1$\times 10^{-10}$ torr. In the current experiment, the MOT loading curves at different background pressures are used to determine the collision cross section between the species of our interest. A leak valve is connected to the quartz cell via an all-metal right angle valve for releasing the background species in a controlled manner. The other end of the leak valve is connected to a gas cylinder containing high purity (99.999$\%$) N$_{2}$ gas, which is used as the background species in our experiment. Since the vacuum pumps connected to the system continuously pump out the background gas, specific leak rates of N$_{2}$ gas is used to achieve desired values of equilibrium pressure of N$_{2}$ in the chamber. This equilibrium pressure is determined by the ion pump current ({\em Agilent}, Model: VacIon Plus 55 Starcell).\ 
 
In order to study the dependence of the loss rate on the power of the trapping laser beams, the MOT is operated at different powers while keeping all other experimental parameters unchanged. The loss of atoms due to escape from the capture region is characterized by measuring the temperature of the atomic cloud. We use the release and recapture technique \cite{RUSSELL2013313} to determine the temperature of the atomic cloud. The release time has been varied from 1 ms to 25 ms for the determination of MOT temperature. 

\section{Results and Discussion}
To measure the $^{88}$Sr$-$N$_{2}$ collision cross section and study the various loss channels, we have used the formulation represented in Eq. \ref{eq2}.  The validity of Eq. \ref{eq2} demands the MOT to be operated in the low-density regime, which is manifested by the exponential growth of the number of atoms in MOT \cite{Rapol2004, PhysRevA.61.051401, PhysRevA.59.1216}. Fig. \ref{fig2} displays four MOT loading curves. These loading curves are recorded in the presence of two repumping lasers operating at (a) 481.3 and (b) 707.2 nm at the respective extreme powers (I$_{1}$ = 46.4 mW/cm$^{2}$, I$_{2}$ = 26.4 mW/cm$^{2}$, I$_{3}$ = 96.8 mW/cm$^{2}$ and I$_{4}$ = 31 mW/cm$^{2}$) of the trapping laser beams.\

\begin{figure}[ht]
	\centering
	\includegraphics[width=0.9\linewidth]{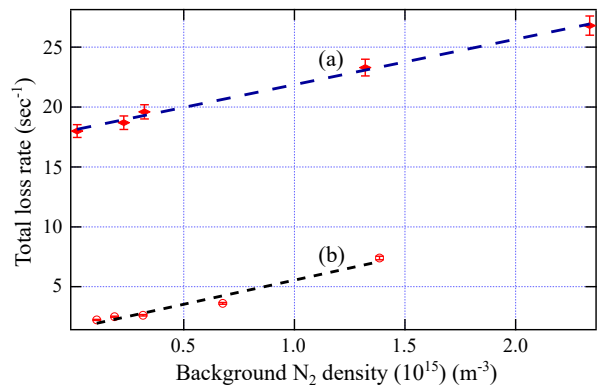} 
	\caption{Loss rate of atoms in Sr MOT as a function of background N$_{2}$ density. The two curves shows the loss rate observed in presence of two repumping schemes. (a) in presence of repumper laser operating at the wavelength 707.2 nm and (b) with a repumping laser at 481.3 nm.}
	\label{fig3}
\end{figure}

To determine the $^{88}$Sr$-$N$_{2}$ collision cross section, the loading curves are recorded at different background pressures. Fig. \ref{fig3} represents the total loss rate of atoms as a function of the density of  N$_{2}$ gas. Considering the loss rate for two different background pressures and using Eq. \ref{eq2}:
\begin{equation}
\frac{1}{\tau_{h}}-\frac{1}{\tau_{l}}= \left(n_{N_{2}}^{h}-n_{N_{2}}^{l}\right)\sigma_{N_{2}}v_{N_{2}} 
\label{eq6}
\end{equation}
Here, the superscript $h$ and $l$ denotes the two different background pressures. On increasing the pressure of N$_{2}$, the total loss rate exhibits a linear increase with a slope of $\sigma_{N_{2}}v_{N_{2}}$ (as predicted by Eq. \ref{eq6}) and is shown in the Fig. \ref{fig3}. The knowledge of the values of  the background densities ($n_{N_{2}}^{h}$ and $n_{N_{2}}^{l}$) and the average velocity ($v_{N_{2}}$) of N$_{2}$ molecules at room temperature enables us to calculate the value of $^{88}$Sr$-$N$_{2}$ collision cross section. For the laser cooled sample of atoms, the average velocity of trapped atoms is negligible with respect to the hot background species, thus for the purpose of calculation we use the absolute average velocity of background N$_{2}$ instead of the relative velocity. Since N$_{2}$ gas is in thermal equilibrium with the vacuum system, we have used $ v_{N_{2}} = 470$ ms$^{-1}$ as calculated at room temperature (20$^{\circ}$C). We found the average value of $^{88}$Sr$-$N$_{2}$ collision cross section to be $ \sigma_{N_{2}} =  $ 8.1(4)$\times 10^{-18}$ m$^{2}$. This result is comparable to  similar experiments performed with Rb and Na using the same background species (N$_{2}$) Ref. \cite{PhysRevA.64.023402}.\

The obtained value of  collision cross section for the ground state ${^1S}_0$ can prove useful to estimate the frequency shift in Sr clock transition due to background N$_{2}$ collisions. These collisions, apart from limiting the trap lifetime, it affects the coherent superposition between the ground ${^1S}_0$ and the metastable exited state ${^3P}_0$ of the clock transition \cite{PhysRevLett.110.180802}. This is manifested as a shift in the clock transition frequency. The interaction potential between a trapped Sr atom in the ground state and a colliding N$_{2}$ molecule can be represented by the Lennard-Jones potential of the form $V(r)=C_{12}/r^{12}- C_{6}/r^{6}$. 

The fractional shift in clock transition frequency, introduced by the collisions with background gases is related to the $C_{6}$ by $\Delta C_{6}/C_{6} \propto \Delta \nu / \nu$, where $\Delta C_{6}$ is the difference between the $C_{6}$ values for the ground and the excited state of the clock transition \cite{PhysRevLett.110.180802} and $\Delta \nu$/$\nu$ is the fractional frequency shift in the clock transition. 
 
The $C_{6}$ for atoms in ${^1S}_0$ state, undergoing collisions with the background N$_{2}$, can be calculated using the experimentally determined collision cross section. For the long range potential of the form V(r)=$-C_{n}/r^{n}$, where $n>2$, the cross section is related to the coefficients in the expansion of the potential as follows \cite{PhysRevA.80.022712}:
\begin{equation}
\sigma(v) \approx A(n) \left(\frac{C_{n}}{\hbar v}\right)^{2/(n-1)}
\label{eq7}
\end{equation}     
Here, $A(n)$ is a constant and $v$  is the relative velocity of colliding particles. The calculated value of $C_{6}$ is found to be 535(68) $E_{H}a_{B}^{6}$ (at $v_{N_{2}} = 470$ ms$^{-1}$, A(6) = 8 \cite{PhysRevA.78.042712}) using the measured value of $\sigma_{N_2}$ above. This formula assumes that there is no trapping force present for the atoms undergoing collisions and hence provide a lower bound for the $C_{6}$.\

For the loss rate measurements performed with cold atoms, there is always a trapping potential present. For such measurements using atoms in the MOT, the trap depth is high enough for an atom to stay trapped in-spite of undergoing a collision. In such scenarios, it is appropriate to use the formula incorporating the trap depth \cite{PhysRevA.85.033420} for deriving $C_{6}$ from loss rate. However, the $C_{6}$ calculated by using this formula gives a value which is an order of magnitude away from the expected value. Similar discrepancy is also observed in the $C_{6}$'s calculated from the other loss rates measured in MOT \cite{PhysRevA.78.042712,PhysRevA.64.023402,Prentiss1988}. The reasons for this inconsistency is not known.\
    
The C$_{6}$ for Sr$-$N$_{2}$ collisions can  also be estimated using the Slater-Kirkwood formula \cite{PhysRevA.85.033420}:
\begin{equation}
C_{6}=\frac{3}{2}\frac{\hbar e}{(4\pi \epsilon_{0})^{2}m_{e}^{1/2}} \frac{\alpha_{Sr}\alpha_{N_{2}}}{(\alpha_{Sr}/\rho_{Sr})^{1/2}+(\alpha_{N_{2}}/\rho_{N_{2}})^{1/2}}
\label{eq8}
\end{equation} 
Here, $m_{e}$ is the electron mass, $\alpha$ and $\rho$ is the static electric polarizability and the number of valence electrons respectively of the colliding species. The estimated $C_{6}$ value from this formula is found to be 310 $E_{H}a_{B}^{6}$. The values of static polarizabilities for the calculation are taken from Refs. \cite{doi:10.1063/1.467820,doi:10.1080/00268976.2018.1535143}. The experimentally obtained $C_{6}$ value for the ground state can be used along with the $C_{6}$ for the excited state in the clock transition to determine the contribution of Sr$-$N$_{2}$ collisions to the clock shift and its error budget. 

\begin{table}[ht]
	\caption{Comparison of literature values for collision cross sections and $C_{6}$ of N$_{2}$ with Rb and Sr. * indicates the values obtained for this work.}
	\centering
	\begin{tabular}{ m{1.5cm} m{1.3cm} m{2cm} m{2cm} m{1cm} } 
		\hline\hline 
		Species \rule{0pt}{3ex} & C$_{6}$ & Loss rate  & $\sigma$ & Ref. \\ [0.5ex]
		& (a.u) & (torr$^{-1}$s$^{-1}$) &  (m$^{2}$) & \\ \hline  
		
		Rb & 302 \rule{0pt}{3ex} & 2.6$\times$10$^{7}$ & -- & \cite{PhysRevA.85.033420} \\ [0.5ex]
		Rb & -- & -- & 3.5(4)$\times$10$^{-18}$ & \cite{PhysRevA.64.023402} \\ [0.5ex]
		Sr & 535 & -- & 8.1(4)$\times$10$^{-18}$ & * \\ [0.5ex]
		\hline\hline
	\end{tabular}
	\label{table1}
\end{table}

Next, we proceed with the estimation of $\tau^{-1}_{power}$ and $\tau^{-1}_{temp}$ in Eq. \ref{eq2} as mentioned in the previous section.  
The presence of repumping laser at 707.2 nm, brings the atoms from ${^3P}_2$ state back to the triplet states ${^3P}_1$ via excited state ${^3S}_1$. The branching ratios for ${^3P}_2$, ${^3P}_1$, and ${^3P}_0$ are 5/9, 3/9, and 1/9, respectively \cite{PhysRevLett.91.053001}. In the absence of 679.3 nm repumping light, the atoms eventually decay into the state ${^3P}_0$ and go out of the cooling cycle. For the second repumping laser operating at 481.3 nm, the branching ratio from the excited state $5p^{2}\ {^3P}_2^{'}$ to the triplet states ${^3P}_2$ and ${^3P}_1$ are 0.714 and 0.286 respectively. The transition from $5p^{2}\ {^3P}_2^{'}$ to ${^3P}_0$ is electric-dipole forbidden.\ 
   
On the other hand, temperature-dependent loss ($\tau^{-1}_{temp}$) denoted by the fourth term in Eq. \ref{eq2}, occurs due to the transfer of atoms into states which are insensitive to the laser used for the first stage cooling transition. Since the atoms in these states do not experience any force, they are free to go out of the MOT capture region. The decay probability into such states is proportional to the fraction of atoms in the excited state ${^1P}_1$.\
 
Since, both the terms mentioned above are a function of the fraction of atoms in the excited state, the separation of these effects is beyond the scope of this experimental technique. Thus we can only observe their cumulative effect.\

\begin{figure}[ht]
	\centering
	\includegraphics[width=0.95\linewidth]{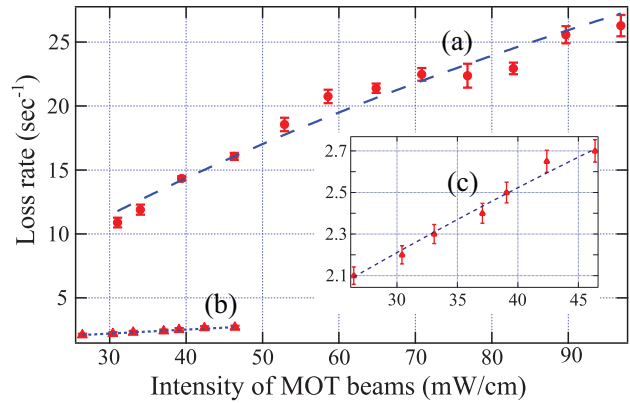} 
	\caption{Loss rate of atoms in Sr MOT as a function of trapping beam intensity. The data is taken for two different configurations of repumping laser, (a) in the presence of repumping laser operating at 707.2 nm, (b) with repumper operating at 481.3 nm. The detuning and axial field gradient for the MOT is kept at -33 MHz and ~48 gauss/cm respectively. Inset (c) is the expanded view of (b).}
	\label{fig4}
\end{figure}

We operate the MOT at different intensities of trapping beams to determine the combined contribution of power and temperature-dependent losses. For  analysis, we have assumed that $\sigma_{N_{2}}$ and $\Gamma_{b}$ are independent of the power of laser beams ($\lambda$ = 460.8 nm)\cite{PhysRevA.78.042712}. Considering the probability of atoms to escape the capture region of MOT ($\epsilon$) to be independent of the trapping beam power, the following relationship can be written using Eq. \ref{eq2}:
\begin{equation}
\frac{1}{\tau_{1}}-\frac{1}{\tau_{2}}= \left(f_{1}-f_{2}\right)\alpha
\label{eq9}
\end{equation}     
Here, the subscripts 1, 2 denotes two different MOT beam powers, and $\alpha$ is the proportionality constant for the combined losses of atoms due to decay into ${^3P}_0$ and due to escape from the trapping region. Thus, $\alpha$ can be written as:
\begin{equation}
\begin{aligned}
\alpha = \alpha_{0}+ \epsilon\gamma_{t}\frac{A_{\ {{^1P}_1\rightarrow\ {^1D}_2}}}{A_{\ {{^1D}_2\rightarrow\ {^3P}_2}}+A_{\ {{^1D}_2\rightarrow\ {^3P}_1}}} 
\end{aligned}
\label{eq10}
\end{equation}
Where the second term is $\tau^{-1}_{temp}$. To check the validity of Eq. \ref{eq9}, we measured the temperature of the atomic cloud with different intensities of MOT beams. The variation in measured temperature with different MOT intensities was found to be $\sim$ 0.3 mK, which is within the error of our experimental data. This allows us to use Eq. \ref{eq9} for the calculation of power and temperature-dependent loss rate. The value of $\alpha$ is extracted by fitting, $1/\tau=\alpha f + c$ to the plot of total loss rate as a function of intensities of MOT beams, as shown in Fig. \ref{fig4}. Employing the two different repumping schemes pumps a different fraction of atoms into ${^3P}_0$ state. The fitting to two different curves gives the value of $\alpha_{707}$ = 167(4) s$^{-1}$ and $\alpha_{481}$ = 18(1) s$^{-1}$. In the above equation, `$c$' is the intercept on the vertical axis and signifies the loss rate, limited solely by the pressure inside the vacuum chamber. The intercept of these curves in both the cases is 1.2(4) s$^{-1}$ showing the background limited lifetime is independent of the repumping schemes.

To get the individual contributions of both the terms in Eq. \ref{eq10}, we theoretically calculate the value of $\tau^{-1}_{power}$ using  Eq. \ref{eq3}. Decay rates for the various energy levels used for the calculations have been taken from Fig. \ref{fig1}. The calculation yields a value of $\alpha_{0}$ $\sim$ 33 s$^{-1}$ for 707.2 nm laser and $\alpha_{0}$ $\sim$ 6.8 s$^{-1}$ for 481.3 nm. The contribution due to temperature-dependent loss channel $\tau^{-1}_{temp}/f$, is obtained by subtracting the value of $\alpha_{0}$ from $\alpha$ and is found to be 134(4) s$^{-1}$ and 11.2(1) s$^{-1}$ for the reumping schemese using 707.2 nm and 481.3 nm respectively.\
 
Experimental determination of the thermal loss channel requires the knowledge of escape probability of the atoms from the capture region which is proportional to the temperature of the atomic cloud. Temperature is measured using the standard release and recapture technique. In this technique, the fraction of atoms recaptured in the trap is measured as a function of release time. This recaptured fraction is inversely proportional to the rate of expansion of the atomic cloud and is used for the determination of temperature.\
 
In the first set of experiments with the MOT beam diameter 10 mm and with a repumping laser operating at the wavelength of 707.2 nm, we get an average temperature of 3.6(3) mK. Eq. \ref{eq5} is used for the calculation of $\epsilon$, and the value is found to be $\sim$ 0.06. This value is used to calculate $\tau^{-1}_{temp}/f$, and the resulting loss rate is found to be 120(15) s$^{-1}$. Incorporating the temperature-dependent losses with the calculated power-dependent losses ($\alpha_{0}$), we get $\alpha$ =  153(15) s$^{-1}$, which agrees with the experimental measurement ($\alpha$ = 167(4) s$^{-1}$) within error.\

For the second set of experiments, we employed a repumping laser with wavelength 481.3 nm, and the diameter of the MOT beams was kept at 15 mm, as mentioned in the earlier section. In this configuration, the temperature is measured to be 3.0(1) mK. For this temperature of the atomic cloud, the calculated value of $\epsilon$ and $\tau^{-1}_{temp}/f$ is found to be 0.0047 and 9.3(8) s$^{-1}$, respectively. On combining the value of $\tau^{-1}_{temp}/f$ and $\alpha_{0}$, we obtain $\alpha$ = 16.1 s$^{-1}$, which is comparable to experimentally obtained value.\ 

\begin{table}[ht]
	\caption{Comparison of experimentally obtained values of $\alpha$, $\alpha_{0}$ and collision cross section between $^{88}$Sr and N$_{2}$ in the presence of two different repumping schemes employing the lasers operating at 707.2 nm and 481.3 nm}
	\centering
	\begin{tabular}{ m{2cm} m{1.5cm} m{1.5cm} m{2.5cm} } 
		\hline\hline 
		$\lambda$ (nm) \rule{0pt}{3ex} & $\alpha$ (s$^{-1}$) & $\alpha_{0}$ (s$^{-1}$) & $\sigma_{^{88}Sr-N_{2}}$ (m$^{2}$) \\ [0.5ex] \hline  
		707.2 & 167(4) \rule{0pt}{3ex} & 33 & 8.1(4)$\times$10$^{-18}$ \\  [0.5ex]
		481.3 & 18(1) & 6.8 & 8.5(8)$\times$10$^{-18}$ \\  [0.5ex]
		\hline\hline
	\end{tabular}
	\label{table1}
\end{table}

The experimentally determined values of $\sigma_{N_{2}}$ and $\alpha$ can be used for the calculation of losses induced by the background species ($\Gamma_{b}$) other than N$_{2}$. This background mostly contains the isotopic mixture of untrapped Sr, H$_{2}$, CO, CO$_{2}$, etc. Using Eq. \ref{eq2}, the value of $\Gamma_b$ is found to be $\sim$0.5 s$^{-1}$.   

\section{Conclusion}
In conclusion, we have studied the loss dynamics of $^{88}$Sr atoms in the blue MOT in two different configurations with repumping lasers operating at the wavelengths 707.2 nm and 481.3 nm. The effect of background N$_{2}$ gas on the total loss rate is studied and is used for the determination of collision cross section between $^{88}$Sr$-$N$_{2}$. The measured value is found to be 8.1(4) $\times 10^{-18}$ m$^{2}$. This value of collision cross section is used for the calculation of $C_{6}$ of ground state Sr atoms and is found to be 535(68) $E_{H}a_{B}^{6}$. We also characterize the various loss channels and determine their contributions towards the total loss rate. We show that the dominant loss channel is from the combined effect of time taken by the atoms to return to the primary cooling cycle via the intermediate states and the thermal escape of atoms from the trapping region during this period. The other contributing loss channel is the decay of atoms into the long-lived state ${^3P}_0$. For the current experimental setup, their collective contribution ($\alpha$) is estimated by operating the MOT at different intensities of trapping beams. The combined decay rate is found to be 167(4) s$^{-1}$ and 18(1) s$^{-1}$ for 707.2 nm and 481.3 nm, respectively.\
 
\section{Acknowledgement}
The authors would like to thank the Department of Science and Technology, Govt. of India for grants through EMR/2014/000365. CV would like to acknowledge Council of Scientific and Industrial Research (CSIR), India for research fellowship. JLM acknowledges the support from NSF GROW (Grant No. DGE 1256260) and IUSSTF.   


%

\end{document}